\documentclass[pra,twocolumn,showpacs,preprintnumbers,superscriptaddress]{revtex4}
\usepackage{xcolor}
\usepackage{times}
\usepackage{bm}
\usepackage{float}
\usepackage{graphicx}
\usepackage{amsbsy}
\usepackage{amsmath}
\usepackage{amsfonts}
\usepackage{amsthm}
\begin{document}

\theoremstyle{plain}
\newtheorem{theorem}{Theorem}
\newtheorem{lemma}[theorem]{Lemma}
\newtheorem{corollary}[theorem]{Corollary}
\newtheorem{proposition}[theorem]{Proposition}
\newtheorem{conjecture}[theorem]{Conjecture}

\theoremstyle{definition}
\newtheorem{definition}[theorem]{Definition}

\theoremstyle{remark}
\newtheorem*{remark}{Remark}
\newtheorem{example}{Example}
\title{Constructing a ball of separable and absolutely separable states for $2\otimes d$ quantum system}
\author{Satyabrata Adhikari}
\email{satyabrata@dtu.ac.in} \affiliation{Delhi Technological
University, Delhi-110042, Delhi, India}

\begin{abstract}
Absolute separable states is a kind of separable state that remain separable under the action of any global unitary transformation. These states may or may not have quantum correlation and these correlations can be measured by quantum discord. We find that the absolute separable states are useful in quantum computation even if it contains infinitesimal quantum correlation in it. Thus to search for the class of two-qubit absolute separable states with zero discord, we have derived an upper bound for $Tr(\varrho^{2})$, where $\varrho$ denoting all zero discord states. In general, the upper bound depends on the state under consideration but if the state belong to some particular class of zero discord states then we found that the upper bound is state independent. Later, it is shown that among these particular classes of zero discord states, there exist sub-classes which are absolutely separable. Furthermore, we have derived necessary conditions for the separability of a given qubit-qudit states. Then we used the derived conditions to construct a ball for $2\otimes d$ quantum system described by $Tr(\rho^{2})\leq Tr(X^{2})+2Tr(XZ)+Tr(Z^{2})$, where the $2\otimes d$ quantum system is described by the density operator $\rho$ which can be expressed by block matrices $X,Y$ and $Z$ with $X,Z\geq 0$. In particular, for qubit-qubit system, we show that the newly constructed ball contain larger class of absolute separable states compared to the ball described by $Tr(\rho^{2})\leq \frac{1}{3}$. Lastly, we have derived the necessary condition in terms of purity for the absolute separability of a qubit-qudit system under investigation.
\end{abstract}
 \maketitle

\section{Introduction}
Quantum correlation can be considered as a necessary ingredient for the development of quantum information theory and quantum computation. A remarkable application of quantum correlation can be found in different areas of quantum communication such as quantum teleportation \cite{bennett}, quantum dense coding \cite{wiesner}, quantum remote state preparation \cite{pati}, quantum cryptography \cite{gisin}  etc.  Till few years ago, it has been thought that this non-local feature in terms of quantum correlation only exist in the entangled state and responsible for the computational speed-up in the known quantum algorithms \cite{ekert}. Later, Lloyd \cite{lloyd} showed that there are quantum search methods which does not require entanglement to provide a computational speed-up over classical methods. In this modern line of research, Ahn et al. \cite{ahn} have shown that instead of entanglement, quantum phase is an essential ingredient for the computational speed-up in the Grover's quantum search algorithm \cite{grover}. Meyer \cite{meyer} was able to reduce the number of queries in a quantum search compared to classical search of a database using only interference, not entanglement. Gottesman-Knill theorem also declare the fact that entanglement is not only a factor for a quantum computers to outperform classical computers \cite{nielsen}. In 2004, E. Biham et.al. \cite{biham} then conclude that entangled state is not a compulsory ingredient for quantum computing and discovered that there exist quantum state lying arbitrarily close to the maximally mixed states, which are enough to increase the computational speed-up in quantum algorithms. To characterize the nature of mixed density matrices lying in the sufficiently small neighbourhood of maximally mixed state, it has been shown that all such states are separable states \cite{Zyczkowski,braunstein,vidal}.\\
Separable states can be defined as the mixture of locally indistinguishable states. Mathematically, a bipartite separable state described by the density operator $\rho$ in a composite Hilbert space $H_{1}\otimes H_{2}$ can be expressed as
\begin{eqnarray}
\rho=\sum_{k} p_{k} \rho_{1}^{k}\otimes \rho_{2}^{k},~~0\leq p_{k}\leq 1
\label{sepstate}
\end{eqnarray}
where $\rho_{1}$ and $\rho_{2}$ represents two density operators in two Hilbert spaces $H_{1}$ and
$H_{2}$ respectively. These states can be prepared using local quantum operation and classical communication (LOCC). Thus prescription of its preparation is different from entangled states, which cannot be prepared with the help of LOCC. Since separable states are prepared by performing quantum operation within the structure of LOCC on quantum bit so they can exhibit quantum correlation \cite{modi}. Therefore, it can be inferred that not only entaglement but also this non-classical feature exhibited by some separable states.\\
The entanglement measures \cite{wootters,vidal1} cannot quantify the quantum correlation present in the separable state due to the reason that any entanglement measure gives value zero for all separable states. Thus, Ollivier and Zurek \cite{ollivier} proposed a measure for quantum correlation which can be defined as the difference between the quantum mutual information and the measurement-induced quantum mutual information. This measure is commonly known as quantum discord. If a separable state has no quantum correlation then they are termed as zero discord state. The quantum discord of two-qubit maximally mixed marginals and two qubit X-state has been calculated in \cite{luo,ali}. It has been found that quantum correlation plays a vital role in mixed-state quantum computation speed-up and it is due to the correlation present in the separable states \cite{dutta2005, dutta2007}. Quantum discord has also been used as a resource in quantum cryptography \cite{pirandola}.\\

We should note an important fact that there exist separable states (with or without quantum correlation) that can be converted to entangled state under the action of global unitary operation. The class of separable states that remain separable state after performing global unitary operation are known as absolutely separable states \cite{kus}. The necessary and sufficient condition for the absolute separability of a state in $2\otimes 2$ system described by the density operator $\sigma$ is given by \cite{verstraete}
\begin{eqnarray}
\lambda_{1}\leq \lambda_{3}+2\sqrt{\lambda_{2}\lambda_{4}}
\label{abssepcond}
\end{eqnarray}
where $\lambda_{i},i=1,2,3,4$ denoting the eigenvalues of $\sigma$ arranged in descending order as $\lambda_{1}\geq \lambda_{2}\geq\lambda_{3}\geq\lambda_{4}$. Further, Johnston \cite{johnston} generalize the absolute separability condition for $H_{2}\otimes H_{d}$ system and showed that a state $\sigma \in H_{2}\otimes H_{d}$ is absolute separable if and only if
\begin{eqnarray}
\lambda_{1}\leq \lambda_{2d-1}+2\sqrt{\lambda_{2d-2}\lambda_{2d}}
\label{abssepcondgen}
\end{eqnarray}
Since the absolute separability conditions (\ref{abssepcond}) and (\ref{abssepcondgen}) depends on the eigenvalues of
the state under investigation so sometimes it is also known as separability from spectrum. Recently, the absolutely separable states are detected and characterized in \cite{nirman,halder}.\\
Let us now discuss the following facts, what we observe in the literature:\\
\textbf{Observation-1:} We can observe that the state used in solving the Deutsch-Jozsa (DJ) problem \cite{deutsch} is a pseudo-pure state (PPS) \cite{gershenfeld} which can be expressed as
\begin{eqnarray}
\rho_{PPS}^{(2)}=\epsilon |\psi\rangle\langle \psi|+\frac{1-\epsilon}{4}I_{4}, 0\leq \epsilon \leq 1
\label{pps}
\end{eqnarray}
where $|\psi\rangle$ is any two-qubit pure state. If $|\psi\rangle$ represent any two-qubit pure maximally entangled state then it reduces to the two-qubit Werner state. The sufficient condition that the state $\rho_{PPS}^{(2)}$ is separable whenever \cite{braunstein}
\begin{eqnarray}
\epsilon < \frac{1}{9}
\label{sepcondpps}
\end{eqnarray}
The quantum algorithm of Deustch and Jozsa solves DJ problem with a single query while classical algorithm uses 3 queries if initially two-qubit PPS with parameter $\epsilon$ $(0\leq \epsilon \leq 1)$ is used in the algorithm. As the number of qubit increases in the initial PPS, the number of queries increases exponentially for classical algorithm while the quantum algorithm of Deustch and Jozsa still require single query and this provide the quantum advantage over classical algorithm. It has been shown that if $\epsilon \leq \frac{1}{33}$ then the initial separable PPS with which the computation has started remain separable throughout the entire computation \cite{biham}. This may imply that the initial PPS is absolutely separable for $0\leq \epsilon \leq \frac{1}{33}$. To verify this statement, let us consider the PPS described by the density operator $\rho_{PPS}$ given by
\begin{eqnarray}
\rho_{PPS}=\epsilon |\psi^{-}\rangle\langle \psi^{-}|+\frac{1-\epsilon}{4}I_{4}, 0\leq \epsilon \leq \frac{1}{33}
\label{newpps}
\end{eqnarray}
where $|\psi^{-}\rangle=\frac{1}{\sqrt{2}}(|01\rangle-|10\rangle)$.\\
The eigenvalues of $\rho_{PPS}$ can be arranged in descending order as $\mu_{1}\geq \mu_{2}\geq \mu_{3}\geq \mu_{4}$, where
\begin{eqnarray}
\mu_{1}=\frac{1+3\epsilon}{4} ,\mu_{2}=\frac{1-\epsilon }{4}, \mu_{3}=\frac{1-\epsilon }{4} , \mu_{4}=\frac{1-\epsilon }{4}
\label{eigvalnew}
\end{eqnarray}
We find that the state $\rho_{PPS}$ is absolutely separable if and only if
\begin{eqnarray}
0\leq \epsilon\leq \frac{1}{33}
\label{abssepex}
\end{eqnarray}
Again, the quantum discord of the state $\rho_{PPS}$ is given by \cite{luo}
\begin{eqnarray}
D(\rho_{PPS})&=&\frac{1-\epsilon}{4}log_{2}(1-\epsilon)-\frac{1+\epsilon}{2}log_{2}(1+\epsilon)\nonumber\\&&+
\frac{1+3\epsilon}{4}log_{2}(1+3\epsilon)
\label{discordluo}
\end{eqnarray}
In particular, if we choose the value of the parameter $\epsilon$ very close to zero, say $\epsilon=0.001$ then $D(\rho_{PPS})\simeq 1.441255\times 10^{-6}$. Therefore, the discord $D(\rho_{PPS})$ is very negligible and can be approximated to zero. Although the quantum correlation of the absolute separable state $\rho_{PPS}$ measured by quantum discord is very near to zero but still it is useful in quantum computation.\\
\textbf{Observation-2:} The largest ball constructed for $d\otimes d$ quantum system is given by $Tr(\rho^{2})\leq \frac{1}{d^{2}-1}$, where the density operator $\rho$ representing either separable or absolutely separable states \cite{Zyczkowski,gurvit}. In particular, it was shown that the largest ball for $2\otimes 2$ quantum system centered at maximally mixed state neither contain all separable states nor absolutely separable states \cite{kus}. To illustrate this, let us consider a state described by the density operator $\sigma_{1}$ given by
\begin{eqnarray}
\sigma_{1}=(\frac{1}{5}|0\rangle\langle 0|)-\frac{4}{5}|1\rangle\langle 1|)\otimes \frac{1}{2}I_{2}
\label{abs1}
\end{eqnarray}
The eigenvalues of $\sigma_{1}$ are given by $\frac{1}{10}, \frac{1}{10}, \frac{2}{5}, \frac{2}{5}$. It can be easily verified that the eigenvalues of $\sigma_{1}$ satisfy the condition (\ref{abssepcond}). Thus the state $\sigma_{1}$ is absolutely separable state. Next, our task is to verify whether the state $\sigma_{1}$ satisfies the inequality $Tr(\sigma_{1}^{2})\leq \frac{1}{3}$. We find that $Tr(\sigma_{1}^{2})=\frac{17}{50}$ which is greater than $\frac{1}{3}$. This implies that the state $\sigma_{1}$ lies outside the ball described by $Tr(\rho^{2})\leq \frac{1}{3}$.\\
The motivation of this work is as follows: Firstly, observation-1 motivate us to search for the class of absolute separable state with zero discord.
Secondly, observation-2 motivate us to construct a ball for $2\otimes d$ quantum system and in particular, we have shown that the constructed ball contain almost all absolute separable states in $2\otimes 2$ dimensional Hilbert space. Thirdly, we find that there are few works that dealt with the quantum communication protocols using separable states with non-classical correlations. In \cite{bobby}, authors found new classes of separable states with non-zero discord that may improve the efficiency of quantum task such as random access code, if the randomness shared between remote partners is described either in the form of finite classical bits or qubits. In another work \cite{jebarathinam}, it has been shown that the quantum task of random access code can be executed efficiently using the finite amount of randomness shared through separable Bell-diagonal states. They have shown that the advantage of their protocol over classical is due to the non-classical correlation known as superunsteerability, which quantifies non-classicality beyond quantum steering. These recent works further motivate us to study quantum discord and absolute separable states.\\

This work is organised as follows: In section-II, we derive the upper bound of $Tr(\varrho^{2})$, where the two-qubit zero discord states are described by the density operator $\varrho$. It is shown that the upper bound is state independent for certain classes of two-qubit zero discord state. These specific classes of two-qubit zero discord states satisfy the condition for the separability from spectrum. In particular, we unearth the class of two-qubit product states which are absolutely separable. Also we find that there exist states from the class of absolute separable zero discord states are not lying within and on a ball described by $Tr(\rho^{2})\leq \frac{1}{3}$. In section-III, we have derived two necessary conditions for the separability of a given qubit-qudit state and then used the derived condition to construct a ball of separable as well as absolute separable state in $2\otimes d$ dimensional system. In section-IV, we give few examples to support that the newly constructed ball is larger in size. It is evident from the fact that in $2\otimes 2$ dimensional system, it contain two-qubit absolute separable states which are lying not only inside but also outside the ball described by $Tr(\rho^{2})\leq \frac{1}{3}$. Further, we provided the example of absolute separable states in $2\otimes 3$ quantum system. In section-V, we have derived necessary condition for the absolute separability of $2\otimes d$ quantum system in terms of purity. In section-VI, we end with concluding remarks.
\section{Identification of a class of absolutely separable states that does not contain quantum correlation}
\noindent In this section, we first derive the upper bound of $Tr(\rho_{ZD}^{2})$, where $\rho_{ZD}$ represent the zero discord state and thereby constructing a ball in which the zero discord state is lying. Then we show that there exist a class of zero discord state residing in the region within the ball, which is separable from spectrum.
\subsection{Construction of a ball that contain zero discord state}
\noindent We construct a ball of zero discord state and to accomplish this task we derive an inequality in terms of $Tr(\rho_{ZD}^{2})$, where the density matrix $\rho_{ZD}$ denoting the zero discord state lying in $2\otimes 2$ dimensional Hilbert space. In general, the derived upper bound of the inequality is state dependent but we found some particular class of zero discord state for which the upper bound is independent of the state.\\
\noindent To start with, let us consider a $2\otimes 2$ dimensional zero discord state $\rho_{ZD}$ that can be expressed as \cite{dakic}
\begin{eqnarray}
\rho_{ZD}= p|\psi\rangle\langle\psi|\otimes \rho_{1}+(1-p)|\psi_{\perp}\rangle\langle\psi_{\perp}|\otimes \rho_{2},~~0\leq p\leq 1
\label{zdstate}
\end{eqnarray}
where the pure states $|\psi\rangle$ and $|\psi_{\perp}\rangle$ are orthogonal to each other i.e. $\langle\psi|\psi_{\perp}\rangle=0$.
The single qubit density operator $\rho_{i} (i=1,2)$ are given by
\begin{eqnarray}
\rho_{i}= \frac{1}{2}I_{2}+\vec{r}_{i}.\vec{\sigma},~~i=1,2
\label{singlequbitdensity}
\end{eqnarray}
$I_{2}$ represent a $2\times 2$ identity matrix, $\vec{r}_{i}=(r_{i1},r_{i2},r_{i3})$ denote the Bloch vector and
the component of $\vec{\sigma}=(\sigma_{1},\sigma_{2},\sigma_{3})$ are usual Pauli matrices.\\
\textbf{Theorem-1:} A two-qubit zero discord state $\rho_{ZD}$ satisfies the inequality
\begin{eqnarray}
Tr(\rho_{ZD}^{2})\leq  min\{\frac{1}{2}+2|\vec{r}_{1}|^{2},\frac{1}{2}+2|\vec{r}_{2}|^{2}\}
\label{th1}
\end{eqnarray}
\textbf{Proof:}
Let us start with the expression of $Tr(\rho_{ZD}^{2})$, which is given by
\begin{eqnarray}
Tr(\rho_{ZD}^{2})= p^{2}Tr(\rho_{1}^{2})+(1-p)^{2}Tr(\rho_{2}^{2})
\label{zdexpression}
\end{eqnarray}
It can be seen that the value of $Tr(\rho_{ZD}^{2})$ is changing by varying the values of the parameter $p$ in the range $0\leq p \leq 1$, and the block vectors $\vec{r_{i}},i=1,2$ satisfying $|\vec{r_{i}}|^{2}\leq 1$. Thus one can ask for the upper bound of $Tr(\rho_{ZD}^{2})$. To probe this question, we assume that the zero discord state $\rho_{ZD}$ satisfies the inequality given by
\begin{eqnarray}
Tr(\rho_{ZD}^{2})\leq \alpha(\vec{r}_{i}),~~i=1,2
\label{assumption}
\end{eqnarray}
regardless of the parameter $p$, where $\alpha(\vec{r}_{i})$ denote the parameter depend on the state parameter $\vec{r}_{i},i=1,2$.\\
Our task is to find $\alpha(\vec{r}_{i})$. To search for $\alpha(\vec{r}_{i})$, we need to combine
 (\ref{zdexpression}) and (\ref{assumption}). Thus, we obtain
\begin{eqnarray}
&&p^{2}Tr(\rho_{1}^{2})+(1-p)^{2}Tr(\rho_{2}^{2})\leq \alpha(\vec{r}_{i}) \nonumber\\&&
\Rightarrow p^{2}(Tr(\rho_{1}^{2})+Tr(\rho_{2}^{2}))-2pTr(\rho_{2}^{2})+\nonumber\\&&(Tr(\rho_{2}^{2})-\alpha(\vec{r}_{i}))\leq 0
\label{step1}
\end{eqnarray}
Solving the inequality (\ref{step1}) for the parameter $p$, we get
\begin{eqnarray}
a\leq p\leq b
\label{step2a}
\end{eqnarray}
where $a$ and $b$ are given by
\begin{eqnarray}
a=\frac{Tr(\rho_{2}^{2})-\sqrt{\alpha(\vec{r}_{i})(Tr(\rho_{1}^{2})+Tr(\rho_{2}^{2}))-Tr(\rho_{1}^{2})Tr(\rho_{2}^{2})}}{Tr(\rho_{1}^{2})+Tr(\rho_{2}^{2})}
\label{step2b}
\end{eqnarray}
and
\begin{eqnarray}
b=\frac{Tr(\rho_{2}^{2})+\sqrt{\alpha(\vec{r}_{i})(Tr(\rho_{1}^{2})+Tr(\rho_{2}^{2}))-Tr(\rho_{1}^{2})Tr(\rho_{2}^{2})}}{Tr(\rho_{1}^{2})+Tr(\rho_{2}^{2})}
\label{step2c}
\end{eqnarray}
We impose the condition on $a$ and $b$ in such a way so that $0\leq p\leq 1$ is satisfied. The required conditions are given below
\begin{eqnarray}
a\geq 0
\label{acond1}
\end{eqnarray}
\begin{eqnarray}
b\leq 1
\label{bcond2}
\end{eqnarray}
The first condition (\ref{acond1}) gives
\begin{eqnarray}
&&\frac{Tr(\rho_{2}^{2})-\sqrt{\alpha(\vec{r}_{i})(Tr(\rho_{1}^{2})+Tr(\rho_{2}^{2}))-Tr(\rho_{1}^{2})Tr(\rho_{2}^{2})}}{Tr(\rho_{1}^{2})+Tr(\rho_{2}^{2})}\geq 0\nonumber\\&&
\Rightarrow  \alpha(\vec{r}_{i})\leq Tr(\rho_{2}^{2})
\label{cond1}
\end{eqnarray}
The second condition (\ref{bcond2}) gives
\begin{eqnarray}
&&\frac{Tr(\rho_{2}^{2})+\sqrt{\alpha(\vec{r}_{i})(Tr(\rho_{1}^{2})+Tr(\rho_{2}^{2}))-Tr(\rho_{1}^{2})Tr(\rho_{2}^{2})}}{Tr(\rho_{1}^{2})+Tr(\rho_{2}^{2})}\leq 1\nonumber\\&&
\Rightarrow \alpha(\vec{r}_{i})\leq Tr(\rho_{1}^{2})
\label{cond2}
\end{eqnarray}
Therefore, (\ref{cond1}) and (\ref{cond2}) can be expressed jointly as
\begin{eqnarray}
\alpha(\vec{r}_{i}) \leq min\{Tr(\rho_{1}^{2}),Tr(\rho_{2}^{2})\}
\label{condcomp}
\end{eqnarray}
Now we calculate $Tr(\rho_{i}^{2})$ by recalling (\ref{singlequbitdensity}), and it is given by
\begin{eqnarray}
Tr(\rho_{i}^{2})=\frac{1}{2}+2|\vec{r}_{i}|^{2}, ~~~i=1,2
\label{trrho}
\end{eqnarray}
Using (\ref{condcomp}) and (\ref{trrho}), we get
\begin{eqnarray}
\alpha(\vec{r}_{i})\leq min\{\frac{1}{2}+2|\vec{r}_{1}|^{2},\frac{1}{2}+2|\vec{r}_{2}|^{2}\}
\label{redineq}
\end{eqnarray}
Combining the inequalities (\ref{assumption}) and (\ref{redineq}), we arrive at the required result
given by
\begin{eqnarray}
Tr(\rho_{ZD}^{2}) \leq min\{\frac{1}{2}+2|\vec{r}_{1}|^{2},\frac{1}{2}+2|\vec{r}_{2}|^{2}\}
\label{th1a}
\end{eqnarray}
Geometrically, the inequality given by (\ref{th1a}) represent a region within and on a ball containing zero discord state. From (\ref{th1a}), it can be easily seen that the upper bound of $Tr(\rho_{ZD}^{2})$ depends on the local bloch vector $\vec{r}_{i}$ and hence the upper bound is state dependent. The state independent bound of $Tr(\rho_{ZD}^{2})$ can be obtained for particular classes of zero discord state and it is given in the corollary below.\\
\textbf{Corollary-1:} The density operators $\rho_{ZD}^{(1)}$ and $\rho_{ZD}^{(2)}$ satisfy the inequality
\begin{eqnarray}
Tr([\rho_{ZD}^{(i)}]^{2})\leq \frac{1}{2},~~i=1,2
\label{cor1}
\end{eqnarray}
where $\rho_{ZD}^{(1)}$ and $\rho_{ZD}^{(2)}$ denote the particular class of zero discord state given by
\begin{eqnarray}
\rho_{ZD}^{(1)}= p|\psi\rangle\langle\psi|\otimes \frac{1}{2}I_{2}+(1-p)|\psi_{\perp}\rangle\langle\psi_{\perp}|\otimes \rho_{2},\nonumber\\ \rho_{ZD}^{(2)}= p|\psi\rangle\langle\psi|\otimes \rho_{1}+(1-p)|\psi_{\perp}\rangle\langle\psi_{\perp}|\otimes \frac{1}{2}I_{2},
\label{zdstate1}
\end{eqnarray}
$0\leq p\leq 1$.\\
\textbf{Proof:} To prove it, consider the following two cases: (i) $|\vec{r}_{1}|^{2}\leq |\vec{r}_{2}|^{2}$, (ii) $|\vec{r}_{2}|^{2}\leq |\vec{r}_{1}|^{2}$.\\
\textbf{Case-I:} If $|\vec{r}_{1}|^{2}\leq |\vec{r}_{2}|^{2}$ then theorem-1 gives
\begin{eqnarray}
Tr(\rho_{ZD}^{2}) \leq \frac{1}{2}+2|\vec{r}_{1}|^{2}
\label{ineq1}
\end{eqnarray}
In particular, the inequality (\ref{ineq1}) holds even if we take the minimum value of the expression $\frac{1}{2}+2|\vec{r}_{1}|^{2}$ over all $\vec{r}_{1}$.
Therefore, we have
\begin{eqnarray}
Tr(\rho_{ZD}^{2}) \leq min_{\vec{r}_{1}}[\frac{1}{2}+2|\vec{r}_{1}|^{2}]
\label{ineq2}
\end{eqnarray}
We obtain $min_{\vec{r}_{1}}[\frac{1}{2}+2|\vec{r}_{1}|^{2}]=\frac{1}{2}$ and the minimum value is attained when $\vec{r}_{1}=\vec{0}$. Thus, the minimum value is obtained when the state $\rho_{ZD}$ reduces to $\rho_{ZD}^{(1)}$. Hence the inequality (\ref{ineq2}) reduces to
\begin{eqnarray}
Tr([\rho_{ZD}^{(1)}]^{2})\leq \frac{1}{2}
\label{cor1a}
\end{eqnarray}
\textbf{Case-II:} If $|\vec{r}_{2}|^{2}\leq |\vec{r}_{1}|^{2}$ then we can proceed in a similar way as in case-I and obtain $Tr([\rho_{ZD}^{(2)}]^{2})\leq \frac{1}{2}$. \\
Therefore, we have obtained the particular classes of zero discord states described by the density operators $\rho_{ZD}^{(i)} (i=1,2)$ given in (\ref{zdstate1}) satisfy the inequality $Tr([\rho_{ZD}^{(i)}]^{2})\leq \frac{1}{2} (i=1,2)$. Thus, the upper bound does not depend on the state $\rho_{ZD}^{(i)} (i=1,2)$.
\subsection{Class of zero discord state which is separable from spectrum}
\noindent Let us consider a class of zero discord state either described by the density operator $\rho_{ZD}^{(1)}$ or $\rho_{ZD}^{(2)}$ given in (\ref{zdstate1}). Recalling $\rho_{ZD}^{(1)}= p|\psi\rangle\langle\psi|\otimes \frac{1}{2}I_{2}+(1-p)|\psi_{\perp}\rangle\langle\psi_{\perp}|\otimes \rho_{2}$ with the single qubit density operator $\rho_{2}$ given by (\ref{singlequbitdensity}) and a pair of orthogonal pure states $|\psi\rangle$ and $|\psi_{\perp}\rangle$, where $|\psi\rangle=\alpha|0\rangle+\beta|1\rangle$ and $|\psi_{\perp}\rangle=\beta|0\rangle-\alpha|1\rangle$. We assume that the parameters $\alpha$ and $\beta$ are real number satisfying $\alpha^{2}+\beta^{2}=1$. Therefore, the density matrix for $\rho_{ZD}^{(1)}$ is given by
\begin{eqnarray}
\rho_{ZD}^{(1)}=
\begin{pmatrix}
  a_{11} & a_{12} & a_{13} & a_{14} \\
  a_{12}^{*} & a_{22} & a_{23} & a_{24} \\
  a_{13}^{*} & a_{23}^{*} & a_{33} & a_{34} \\
  a_{14}^{*} & a_{24}^{*} & a_{34}^{*} & a_{44}
\end{pmatrix}, \sum_{i=1}^{4}a_{ii}=1
\end{eqnarray}
where
\begin{eqnarray}
&&a_{11}=p\frac{\alpha^{2}}{2}+(1-p)\beta^{2}(\frac{1}{2}+r_{23}),\nonumber\\&& a_{12}=(1-p)\beta^{2}(r_{21}-ir_{22}),\nonumber\\&& a_{13}=p\frac{\alpha\beta}{2}-(1-p)\alpha\beta(\frac{1}{2}+r_{23}),\nonumber\\&&
a_{14}=-(1-p)\alpha\beta(r_{21}-ir_{22}), \nonumber\\&& a_{22}=p\frac{\alpha^{2}}{2}+(1-p)\beta^{2}(\frac{1}{2}-r_{23}),\nonumber\\&& a_{23}=-(1-p)\alpha\beta(r_{21}+ir_{22}),\nonumber\\&&
a_{24}=p\frac{\alpha\beta}{2}-(1-p)\alpha\beta(\frac{1}{2}-r_{23}),\nonumber\\&&
a_{33}=p\frac{\beta^{2}}{2}+(1-p)\alpha^{2}(\frac{1}{2}+r_{23}),\nonumber\\&&
a_{34}=(1-p)\alpha^{2}(r_{21}-ir_{22}),\nonumber\\&&
a_{44}=p\frac{\beta^{2}}{2}+(1-p)\alpha^{2}(\frac{1}{2}-r_{23})
\label{matrix}
\end{eqnarray}
The eigenvalues of $\rho_{ZD}^{(1)}$ are given by
\begin{eqnarray}
&&\lambda_{1}=\frac{1-p}{2}(1+2|\vec{r}_{2}|),\lambda_{2}=\frac{1-p}{2}(1-2|\vec{r}_{2}|)\nonumber\\&&
\lambda_{3}=\lambda_{4}=\frac{p}{2}
\label{eigval}
\end{eqnarray}
The state $\rho_{ZD}^{(1)}$ satisfy the positive semi-definiteness property if
\begin{eqnarray}
|\vec{r}_{2}|\leq \frac{1}{2}
\label{possem}
\end{eqnarray}
Now our task reduces to the following; (i) verify whether the class of states $\rho_{ZD}^{(1)}$ satisfy the condition of separability from spectrum and (ii) if the class of states verified as absolute separable states then find out whether they lying within the ball described by $Tr([\rho_{ZD}^{(1)}]^{2})\leq\frac{1}{3}$. In this context, a table is constructed by taking different ranges of the parameter $p$ and some values of $|\vec{r}_{2}|$ for which we find that the zero discord state described by the density operator $\rho_{ZD}^{(1)}$ satisfy the inequality (\ref{abssepcond}). This means that there exist classes of two-qubit zero discord states that are absolutely separable also. We call these classes of two-qubit states as Absolutely Separable Zero Discord Class $(ASZDC)$. Further, we have constructed another table which reveals the fact that whether the class of states given by $ASZDC$ satisfies the inequality $Tr([\rho_{ASZDC}]^{2})\leq\frac{1}{3}$. Without any loss of generality, we have verified the above two tasks by considering the values of the parameter $p$ in $[0,\frac{1}{2}]$ and taking few values of $|\vec{r_{2}}|$. Similar analysis can be done for other range the parameter $p\in [\frac{1}{2},1]$ and other values of $|\vec{r_{2}}|\leq \frac{1}{2}$.\\
\begin{table}
\begin{center}
\caption{Table verifying whether the state $\rho_{ZD}^{(1)}$ satisfy (\ref{abssepcond}) and whether residing inside or outside the ball described by
$Tr([\rho_{ASZDC}]^{2})\leq\frac{1}{3}$}
\begin{tabular}{|c|c|c|c|}\hline
$Parameter$ & $Parameter$ & $Separable /$ & $Outside /$  \\ $(|\vec{r}_{2}|)$ & (p)  & $Absolute Separable$ &  $Inside$  \\  \hline
0  & [0, 0.15) & Separable & Outside \\
0  & [0.15, 0.211) & Absolute separable & Outside \\
0  & [0.211, 0.5] & Absolute separable & Inside \\\hline
0.1  & [0, 0.213) & Separable & Outside \\
0.1  & [0.213, 0.2325) & Absolute separable & Outside  \\
0.1  & [0.2325, 0.5) & Absolute separable & Inside  \\\hline
0.2  & [0, 0.291) & Separable & Outside \\
0.2  & [0.291, 0.29205) & Absolute separable  & Outside  \\
0.2  & [0.29205, 0.5) & Absolute separable  & Inside  \\\hline
0.3  & [0, 0.38) & Separable & Outside \\
0.3  & [0.38, 0.38056) & Absolute separable & Outside \\
0.3  & [0.38056, 0.5) & Absolute separable & Inside \\\hline
0.4  & [0, 0.483) & Separable & Outside \\
0.4  & [0.483, 0.49) & Absolute separable & Outside \\
0.4  & [0.49, 0.5) & Absolute separable & Inside \\\hline
0.5  & [0, 0.5] & Separable & Outside \\\hline
\end{tabular}
\end{center}
\end{table}

\noindent Since the maximal ball described by $Tr([\rho_{ASZDC}]^{2})\leq\frac{1}{3}$ does not contain all states from the class ASZDC and such states lying outside the ball so we investigate in the next section that whether it is possible to increase the size of the maximal ball.
\section{Constructing the bigger ball of separable as well as absolutely separable states around maximally mixed state}
In this section, we will show that it is possible to construct a ball which is larger than the earlier constructed
ball described by $Tr(\rho^{2})\leq\frac{1}{3}$ where the state $\rho$ represent either separable or absolutely separable states around maximally mixed state. This means that there is a possibility for the new ball, constructed in this work, to contain those separable as well as absolute separable states which are lying outside the ball described by $Tr(\rho^{2})\leq\frac{1}{3}$.
\subsection{A Few Definitions and Results}
\noindent Firstly, we recapitulate a few definitions and earlier obtained results which are required to construct a new ball.\\
\textbf{Definition-1:} \textsl{p}-norm of a matrix $A$ is defined as
\begin{eqnarray}
(\|A\|_{p})^{p}=Tr(A^{\dagger}A)^{\frac{p}{2}}
\label{pnorm}
\end{eqnarray}
In particular for $p=2$ and $A=\rho$, where $\rho$ denoting a quantum state, we have
\begin{eqnarray}
(\|\rho\|_{2})^{2}=Tr(\rho^{2})
\label{pnorm1}
\end{eqnarray}
\textbf{Definition-2: \cite{hilderbrand}} A quantum state $\rho \in H_{2}\otimes H_{d}$ is absolutely separable if $U\rho U^{\dagger}$ remain a separable state for
all global unitary operator $U \in U(2d)$.\\
If we denote $\rho'=U\rho U^{\dagger}$ then it can be easily shown that $Tr[(\rho')^{2}]=Tr[(\rho)^{2}]$, i.e. $Tr[(\rho)^{2}]$ is
invariant under unitary transformation.\\
\textbf{Result-1 \cite{king}:} Let $M$ be a $2d\times 2d$ positive semi-definite matrix expressed in the block form as
\begin{eqnarray}
M=
\begin{pmatrix}
  A & C \\
  C^{\dag} & B
\end{pmatrix}
\end{eqnarray}
where $A,B,C$ are $d\times d$ matrices.\\
If we define the $2\times 2$ matrix as
\begin{eqnarray}
m=
\begin{pmatrix}
  \|A\|_{p} &  \|C\|_{p}\\
   \|C\|_{p} &  \|B\|_{p}
\end{pmatrix}
\end{eqnarray}
then the following inequalities hold:\\
(a) for $1\leq p \leq 2$,\\
\begin{eqnarray}
\|M\|_{p}\geq \|m\|_{p}
\label{inequalitya}
\end{eqnarray}
(b) for $2\leq p < \infty$,\\
\begin{eqnarray}
\|M\|_{p}\leq \|m\|_{p}
\label{inequalityb}
\end{eqnarray}
Thus for $p=2$, we have
\begin{eqnarray}
\|M\|_{2}= \|m\|_{2}
\label{inequality}
\end{eqnarray}
\textbf{Result-2 \cite{johnston}:} Let us choose $d\times d$ matrices $A,B,C$ such that $A$ and $B$ are positive semi-definite matrices. Then the block matrix
\begin{eqnarray}
X=
\begin{pmatrix}
  A & C \\
  C^{\dag} & B
\end{pmatrix}
\end{eqnarray}
is separable if $\|C\|_{2}^{2}\leq \lambda_{min}(A)\lambda_{min}(B)$, where $\lambda_{min}(A)$ and $\lambda_{min}(B)$ denoting the minimum eigenvalue
of the matrices $A$ and $B$ respectively.\\
\subsection{A Necessary condition for the Separability}
Let us consider a $2\otimes d$ dimensional quantum system described by the density operator $\rho_{AB}$ as
\begin{eqnarray}
\rho_{AB}=
\begin{pmatrix}
  X & Y \\
  Y^{\dagger} & Z
\end{pmatrix}
\label{th2blockmatrix}
\end{eqnarray}
where $X,Y,Z$ are $d\times d$ block matrices.\\
\textbf{Theorem-2:}  If the state $\rho_{AB}$ is separable then
\begin{eqnarray}
Tr(XZ)\geq Tr(YY^{\dagger})
\label{th2cond}
\end{eqnarray}
\textbf{Proof:} The reduced density matrix $\rho_{B}$ is given by
\begin{eqnarray}
Tr_{A}(\rho_{AB})=\rho_{B}=X+Z
\label{reddensity2}
\end{eqnarray}
The linear entropy $S_{L}$ of the reduced state $\rho_{B}$ is given by
\begin{eqnarray}
S_{L}(\rho_{B})&=&1-Tr(\rho_{B}^{2})\nonumber\\&=&1-Tr(X^{2})-2Tr(XZ)-Tr(Z^{2})
\label{linearentb}
\end{eqnarray}
Also, we have
\begin{eqnarray}
Tr(\rho_{AB}^{2})=Tr(X^{2})+2Tr(YY^{\dagger})+Tr(Z^{2})
\label{trace20}
\end{eqnarray}
Therefore, the linear entropy of the composite system $\rho_{AB}$ is given by
\begin{eqnarray}
S_{L}(\rho_{AB})&=&1-Tr(\rho_{AB}^{2})\nonumber\\&=&2Tr(XZ)+S_{L}(\rho_{B})-2Tr(YY^{\dagger})
\label{linearentab}
\end{eqnarray}
It is known that \cite{santos} if the state $\rho_{AB}$ is separable then
\begin{eqnarray}
S_{L}(\rho_{AB})\geq S_{L}(\rho_{B})
\label{linearentineq1}
\end{eqnarray}
From (\ref{linearentab}) and (\ref{linearentineq1}), we have
\begin{eqnarray}
Tr(XZ)\geq Tr(YY^{\dagger})
\label{necessarycond}
\end{eqnarray}
Hence proved.\\
\textbf{Corollary-2:} If there exist any state $\sigma_{AB} \in H_{2}\otimes H_{d}$ that violate the condition (\ref{th2cond}) then the state $\sigma_{AB}$ is definitely an entangled state i.e. if the state described by the density operator $\sigma_{AB}=\begin{pmatrix}
  X & Y \\
  Y^{\dagger} & Z
\end{pmatrix}$, satisfies the inequality
\begin{eqnarray}
Tr(XZ)< Tr(YY^{\dagger})
\label{th2cond10}
\end{eqnarray}
then the state $\sigma_{AB}$ is an entangled state.\\
For instance, let us consider a state $\varsigma_{AB} \in H_{2}\otimes H_{4}$ described by the density operator
\begin{eqnarray}
\varsigma_{AB}=
\begin{pmatrix}
  \frac{a}{6a+1} & 0 & 0 & 0 & 0 & 0 & 0 & \frac{a}{6a+1} \\
  0 & \frac{a}{6a+1} & 0 & 0 & 0 & 0 & \frac{a}{6a+1} & 0 \\
  0 & 0 & \frac{a}{6a+1} & 0 & 0 & \frac{a}{6a+1} & 0 & 0 \\
  0 & 0 & 0 & 0 & 0 & 0 & 0 & 0 \\
  0 & 0 & 0 & 0 & 0 & 0 & 0 & 0 \\
  0 & 0 & \frac{a}{6a+1} & 0 & 0 & \frac{a}{6a+1} & 0 & 0 \\
  0 & \frac{a}{6a+1} & 0 & 0 & 0 & 0 & \frac{a}{6a+1} & 0 \\
   \frac{a}{6a+1} & 0 & 0 & 0 & 0 & 0 & 0 & \frac{1+a}{6a+1} \\
\end{pmatrix}
\end{eqnarray}
where $a\in[0,1]$. The $4\times 4$ block matrices for the state $\varsigma_{AB}$ is given by\\
\begin{eqnarray}
&&X=
\begin{pmatrix}
 \frac{a}{6a+1}  &  0 & 0 & 0\\
 0  & \frac{a}{6a+1} & 0 & 0\\
 0  & 0 & \frac{a}{6a+1} & 0\\
 0  & 0 & 0 & 0
\end{pmatrix}, Y=\begin{pmatrix}
  0  & 0 & 0 & \frac{a}{6a+1}\\
  0  & 0 & \frac{a}{6a+1} & 0\\
  0  & \frac{a}{6a+1} & 0 & 0\\
   0  & 0 & 0 & 0\\
\end{pmatrix},\\&& Z=\begin{pmatrix}
   0  & 0 & 0 & 0\\
   0  & \frac{a}{6a+1} & 0 & 0\\
   0  & 0 & \frac{a}{6a+1} & 0\\
   0  & 0 & 0 & \frac{a}{6a+1}\\
\end{pmatrix}
\end{eqnarray}
Therefore, we have
\begin{eqnarray}
Tr(XZ)=\frac{2a^{2}}{(6a+1)^{2}},~~ Tr(YY^{\dagger})=\frac{3a^{2}}{(6a+1)^{2}}
\label{exampleentcond}
\end{eqnarray}
Thus, for $0<a\leq 1$, we have obtained $Tr(XZ)< Tr(YY^{\dagger})$. Hence the state $\varsigma_{AB}$ is an entangled state.\\
Further, it can be easily shown that $\varsigma_{AB}^{T_{B}}$ has one negative eigenvalue and thus we can again verify that the state
described by the density operator $\varsigma_{AB}$ is an entangled state.\\
\textbf{Theorem-3:} If the state $\rho_{AB}$ described by the density operator (\ref{th2blockmatrix}) is separable then
\begin{eqnarray}
S_{L}(\rho_{A})-S_{L}(\rho_{B})\leq 2[Tr(XZ)-Tr(YY^{\dagger})]
\label{th3cond}
\end{eqnarray}
\textbf{Proof:} The reduced density matrix $\rho_{A}$ is given by
\begin{eqnarray}
Tr_{B}(\rho_{AB})=\rho_{A}=\begin{pmatrix}
  TrX & TrY \\
  TrY^{\dagger} & TrZ
\end{pmatrix}
\label{reddensity1}
\end{eqnarray}
The linear entropy $S_{L}$ of the reduced state $\rho_{A}$ is given by
\begin{eqnarray}
S_{L}(\rho_{B})&=&1-Tr(\rho_{A}^{2})\nonumber\\&=&1-(Tr(X))^{2}-2Tr(Y)Tr(Y^{\dagger})\nonumber\\&-&(Tr(Z))^{2}
\label{linearenta}
\end{eqnarray}
It is known that \cite{santos} if the state $\rho_{AB}$ is separable then
\begin{eqnarray}
S_{L}(\rho_{AB})\geq S_{L}(\rho_{A})
\label{linearentineq2}
\end{eqnarray}
Using the expression of the linear entropy of the composite system $\rho_{AB}$ given by (\ref{linearentab}) in (\ref{linearentineq2}), we get
$S_{L}(\rho_{A})-S_{L}(\rho_{B})\leq 2[Tr(XZ)-Tr(YY^{\dagger})]$.\\
Hence proved.\\
\textbf{Corollary-3:} If any qubit-qudit state violate the condition (\ref{th3cond}) then the qubit-qudit state is definitely an entangled state.
\subsection{Construction of a new ball that contain separable as well as absolutely separable states}
Let us consider a quantum state described by the density matrix $\rho \in H_{2}\otimes H_{d}$. The density matrix can be written in the block form as
\begin{eqnarray}
\rho=
\begin{pmatrix}
  X & Y \\
  Y^{\dagger} & Z
\end{pmatrix}
\end{eqnarray}
where $X,Y,Z$ denoting $d\times d$ matrices with $X,Z\geq 0$.\\
Using Result-1, we have
\begin{eqnarray}
\|\rho\|_{2}=
\|\begin{pmatrix}
  X &  Y\\
   Y^{\dagger} &  Z
\end{pmatrix}\|_{2}=\|\begin{pmatrix}
  \|X\|_{2} &  \|Y\|_{2}\\
   \|Y\|_{2} &  \|Z\|_{2}
\end{pmatrix}\|_{2}
\end{eqnarray}
Let us now calculate the value of $Tr(\rho^{2})$. It is given by
\begin{eqnarray}
Tr(\rho^{2})&=&\|\rho\|_{2}^{2}= \|X\|_{2}^{2}+2\|Y\|_{2}^{2}+\|Z\|_{2}^{2}
\nonumber\\&=& Tr(X^{2})+2Tr(YY^{\dagger})+Tr(Z^{2})
\label{trace10}
\end{eqnarray}
Now, we are in a position to construct a ball based on two separability conditions: (i) separability condition given in Result-2 and (ii) separability condition derived in Theorem-2.\\
\textbf{Result-3:} If the state $\rho \in H_{2}\otimes H_{d} $ is separable then it contained in the ball $(B_{1})$ given by
\begin{eqnarray}
Tr(\rho^{2})\leq Tr(X^{2})+2\lambda_{min}(X)\lambda_{min}(Z)+Tr(Z^{2})
\label{trace20}
\end{eqnarray}
where $\lambda_{min}(X)$ and $\lambda_{min}(Z)$ denoting the minimum eigenvalues of the block matrices $X$ and $Z$ respectively.\\
\textbf{Proof:} If the state $\rho$ is separable then from result-2, we have
\begin{eqnarray}
Tr(YY^{\dagger})\leq \lambda_{min}(X)\lambda_{min}(Z)
\label{result-21}
\end{eqnarray}
Using (\ref{result-21}) in (\ref{trace10}), we get
\begin{eqnarray}
Tr(\rho^{2})\leq Tr(X^{2})+2\lambda_{min}(X)\lambda_{min}(Z)+Tr(Z^{2})
\label{trace30}
\end{eqnarray}
The state described by the density operator $\rho$ is absolutely separable if for any global
unitary transformation $U \in U(2d)$, the inequality
\begin{eqnarray}
Tr[(U\rho U^{\dagger})^{2}]=Tr(\rho^{2})&\leq& Tr(X^{2})+2\lambda_{min}(X)\lambda_{min}(Z)\nonumber\\&+&Tr(Z^{2})
\label{abscond}
\end{eqnarray}
holds.\\
\textbf{Result-4:} If the state $\rho \in H_{2}\otimes H_{d} $ is separable then it contained in the ball $(B_{2})$ given by
\begin{eqnarray}
Tr(\rho^{2})\leq Tr(X^{2})+2Tr(XZ)+Tr(Z^{2})
\label{trace200}
\end{eqnarray}
\textbf{Proof:} If the state $\rho$ is separable then from theorem-2, we have
\begin{eqnarray}
Tr(YY^{\dagger})\leq Tr(XZ)
\label{theorem-210}
\end{eqnarray}
Using (\ref{theorem-210}) in (\ref{trace10}), we get
\begin{eqnarray}
Tr(\rho^{2})\leq Tr(X^{2})+2Tr(XZ)+Tr(Z^{2})
\label{trace300}
\end{eqnarray}
The state described by the density operator $\rho$ is absolutely separable if for any global
unitary transformation $U \in U(2d)$, the inequality
\begin{eqnarray}
Tr[(U\rho U^{\dagger})^{2}]=Tr(\rho^{2})&\leq& Tr(X^{2})+2Tr(XZ)\nonumber\\&+&Tr(Z^{2})
\label{abscond1}
\end{eqnarray}
holds.\\
It can be observed that the upper bound of the inequalities (\ref{abscond}) and (\ref{abscond1}) depends on the parameter of the state under consideration. Thus the upper bound is state dependent and it can be maximized over the given range of the parameter of the state. We grasp this idea to show that there is a possibility to increase the size of the ball that contains more separable as well as absolutely separable state compared to $Tr(\rho^{2})\leq \frac{1}{3}$.
\section{Illustrations}
\noindent In this section, we will show with examples that the new ball constructed in this work described by (\ref{abscond}) contains more two-qubit absolutely separable
states than the ball descibed by $Tr(\rho^{2})\leq \frac{1}{3}$. Also, we discuss about the absolute separable states in $2\otimes 3$ quantum system.
\subsubsection{Two-qubit class of sates from ASZDC}
Let us consider a subclass of the two-qubit quantum state belong to ASZDC described by the density operator $\rho^{(1)}$ as
\begin{eqnarray}
\rho^{(1)}&=& (p|0\rangle\langle 0|+(1-p)|1\rangle\langle 1|)\otimes \frac{1}{2}I_{2},\nonumber\\&&
0\leq p\leq 1
\label{class1a}
\end{eqnarray}
where $I_{2}$ represent the identity matrix of order 2. The state $\rho^{(1)}$ is a product state and thus separable for~ $0\leq p\leq 1$.\\
The matrix representation of $\rho^{(1)}$ is given by
\begin{eqnarray}
\rho^{(1)}=
\begin{pmatrix}
  X & Y \\
  Y^{\dagger} & Z
\end{pmatrix}
\end{eqnarray}
where $Y$ is a null matrix and the matrices $X$ and $Z$ are given by
\begin{eqnarray}
&&X=
\begin{pmatrix}
  \frac{p}{2}& 0 \\
  0 & \frac{p}{2}
\end{pmatrix},\nonumber\\&& Z=\begin{pmatrix}
  \frac{1-p}{2}& 0 \\
  0 & \frac{1-p}{2}
\end{pmatrix}
\end{eqnarray}
The eigenvalues of $\rho^{(1)}$ are given by $\frac{p}{2}, \frac{p}{2},\frac{1-p}{2},\frac{1-p}{2}$.\\
Case-I: When the parameter $p$ is lying in the interval $[0,\frac{1}{2}]$ then the eigenvalues are arranged in descending order as
$\lambda_{1}\geq \lambda_{2}\geq\lambda_{3}\geq\lambda_{4}$, where
\begin{eqnarray}
\lambda_{1}=\frac{1-p}{2}, \lambda_{2}=\frac{1-p}{2},\lambda_{3}=\frac{p}{2},\lambda_{4}=\frac{p}{2}
\label{eigenvalue1}
\end{eqnarray}
The state $\rho^{(1)}$ is separable from spectrum if
\begin{eqnarray}
p+\sqrt{p(1-p)}\geq \frac{1}{2}
\label{eigenvalue11}
\end{eqnarray}
The inequality (\ref{eigenvalue11}) holds if $\frac{3}{20}\leq p\leq 1/2$. Therefore, the state $\rho^{(1)}$
is absolutely separable for $p \in [\frac{3}{20},\frac{1}{2}]$.\\
Now, $Tr[(\rho^{(1)})^{2}]$ can be calculated as
\begin{eqnarray}
Tr[(\rho^{(1)})^{2}]=\frac{p^{2}}{2}+\frac{(1-p)^{2}}{2}
\label{trace11}
\end{eqnarray}
From Fig.1, it can be seen that there exist absolutely separable states for $p \in [\frac{3}{20}, \frac{21}{100}]$ that are lying outside the ball described by $Tr[(\rho^{(1)})^{2}]\leq\frac{1}{3}$. Thus, it is interesting to see whether the newly constructed ball contain all the absolutely separable states for $p \in [\frac{3}{20}, \frac{21}{100}]$. To probe this, we calculate the upper bound of $Tr[(\rho^{(1)})^{2}]$ using the inequality (\ref{abscond}) and (\ref{trace200}). The upper bounds for the balls $B_{1}$ and $B_{2}$ are given by
\begin{eqnarray}
&&Tr(X^{2})+2\lambda_{min}(X)\lambda_{min}(Z)+Tr(Z^{2})\nonumber\\&&=\frac{1}{2}[1-p(1-p)]
\label{ub1}
\end{eqnarray}
and
\begin{eqnarray}
Tr(X^{2})+2Tr(XZ)+Tr(Z^{2})=\frac{1}{2}
\label{ub2}
\end{eqnarray}
Again, Fig.1 shows that the newly constructed balls $B_{1}$ and $B_{2}$ contains all absolutely separable belong to the class described by the density operator $\rho^{(1)}$. Also, it can be seen that Ball $B_{2}$ contains more absolutely separable states than ball $B_{1}$.\\
\noindent Case-II: In a similar fashion, the case where $p \in [\frac{1}{2},1]$ can be analyzed.
\begin{figure}[h]
\centering
\includegraphics[scale=0.4]{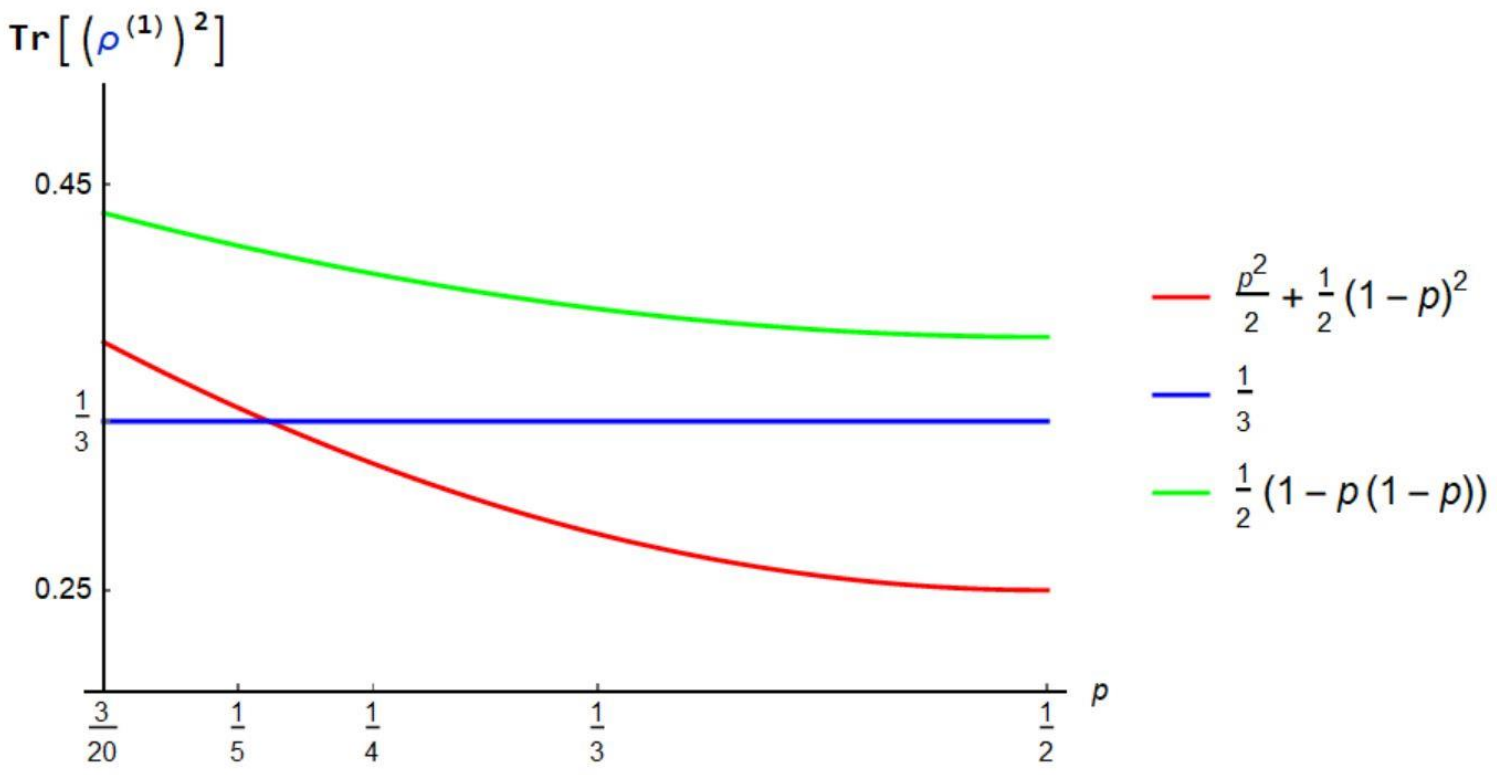}
\caption{Plot of $Tr[(\rho^{(1)})^{2}]$ versus the state parameter $p$ }
\end{figure}
\subsubsection{$2\times 2$ isotropic state}
Let us consider a $2\otimes 2$ isotropic state given by
\begin{eqnarray}
\rho_{2\otimes 2}^{(iso)}(f)=
\begin{pmatrix}
  \frac{1+2f}{6} & 0 & 0 & \frac{4f-1}{6} \\
  0 & \frac{1-f}{3} & 0 & 0 \\
  0 & 0 & \frac{1-f}{3} & 0 \\
  \frac{4f-1}{6} & 0 & 0 & \frac{1+2f}{6}
\end{pmatrix}, 0\leq f\leq 1
\end{eqnarray}
It is known that the state described by the density operator $\rho_{2\otimes 2}^{(iso)}$ is separable for $0\leq f\leq \frac{1}{2}$. Further, it can be easily verified that all separable states in the class represented by $\rho_{2\times 2}^{(iso)}$ are also absolute separable states.\\
The matrix of $2\otimes 2$ isotropic state can be re-expressed in terms of block matrices of order $2\times 2$  as
\begin{eqnarray}
\rho_{2\otimes 2}^{(iso)}(f)=
\begin{pmatrix}
  X & Y \\
  Y^{\dagger} & Z
\end{pmatrix}
\end{eqnarray}
where $2\times 2$ block matrices $X,Y$ and $Z$ are given by
\begin{eqnarray}
X=
\begin{pmatrix}
 \frac{1+2f}{6}  &  0\\
   0 &  \frac{1-f}{3}
\end{pmatrix}, Y=\begin{pmatrix}
  0 & \frac{4f-1}{6}\\
   0 &  0
\end{pmatrix}, Z=\begin{pmatrix}
  \frac{1-f}{3} & 0\\
   0 &  \frac{1+2f}{6}
\end{pmatrix}
\end{eqnarray}
The minimum eigenvalue of the block matrices $X$ and $Z$ are given by
\begin{eqnarray}
\lambda_{min}(X)=\lambda_{min}(Z)&=&\frac{1+2f}{6},~~0\leq f\leq \frac{1}{4}\nonumber\\&=&
\frac{1-f}{3},~~\frac{1}{4}\leq f\leq \frac{1}{2}
\label{mineigen}
\end{eqnarray}
We now discuss two cases based on different ranges of the parameter $f$.\\
\textbf{Case-I:} When $0\leq f \leq \frac{1}{4}$
\begin{eqnarray}
Tr[(\rho_{2\otimes 2}^{(iso)}(f))^{2}]\leq \frac{2f^{2}+1}{3}
\label{trace1}
\end{eqnarray}
Since $f \in [0,\frac{1}{4}]$ so (\ref{trace1}) can be re-expressed as
\begin{eqnarray}
Tr[(\rho_{2\otimes 2}^{(iso)}(f))^{2}]\leq Max_{0\leq f\leq\frac{1}{4}}\frac{2f^{2}+1}{3}
\label{trace11}
\end{eqnarray}
Since $\frac{2f^{2}+1}{3}$ is an increasing function of the parameter $f$ so its maximum value is attained at $f=\frac{1}{4}$. Therefore,
\begin{eqnarray}
Max_{0\leq f\leq\frac{1}{4}}\frac{2f^{2}+1}{3}=\frac{3}{8}
\label{max1}
\end{eqnarray}
Thus, the state $\rho_{2\otimes 2}^{(iso)}(f)$ satisfies the inequality given by
\begin{eqnarray}
Tr[(\rho_{2\otimes 2}^{(iso)}(f))^{2}]\leq \frac{3}{8}
\label{case1}
\end{eqnarray}
\textbf{Case-II:} When $\frac{1}{4}\leq f \leq \frac{1}{2}$
\begin{eqnarray}
Tr[(\rho_{2\otimes 2}^{(iso)}(f))^{2}]\leq \frac{4f^{2}-4f+3}{6}
\label{trace2}
\end{eqnarray}
Since $f \in [\frac{1}{4},\frac{1}{2}]$ so (\ref{trace2}) can be reexpressed as
\begin{eqnarray}
Tr[(\rho_{2\otimes 2}^{(iso)}(f))^{2}]\leq Max_{\frac{1}{4}\leq f\leq\frac{1}{2}}\frac{4f^{2}-4f+3}{6}
\label{trace22}
\end{eqnarray}
Since $\frac{4f^{2}-4f+3}{6}$ is a decreasing function of the parameter $f$ so its maximum value is attained at $f=\frac{1}{4}$. Therefore,
\begin{eqnarray}
Max_{\frac{1}{4}\leq f\leq\frac{1}{2}}\frac{4f^{2}-4f+3}{6}= \frac{3}{8}
\label{max2}
\end{eqnarray}
Thus, in this case also the state $\rho_{2\otimes 2}^{(iso)}(f)$ obey the inequality given by
\begin{eqnarray}
Tr[(\rho_{2\otimes 2}^{(iso)}(f))^{2}]\leq \frac{3}{8}
\label{case2}
\end{eqnarray}
Combining the above two cases, it can be concluded that the state $\rho_{2\otimes 2}^{(iso)}(f)$ satisfy the inequality
\begin{eqnarray}
Tr[(\rho_{2\otimes 2}^{(iso)}(f))^{2}]\leq\frac{3}{8}, 0\leq f\leq\frac{1}{2}
\label{abssep1}
\end{eqnarray}
Furthermore, Eq. (\ref{trace200}) described the ball $B_{2}$ for the state $\rho_{2\otimes 2}^{(iso)}(f)$ as
\begin{eqnarray}
Tr[(\rho_{2\otimes 2}^{(iso)}(f))^{2}]\leq\frac{4f^{2}-2f+7}{18}, 0\leq f\leq\frac{1}{2}
\label{abssep100}
\end{eqnarray}
Eq. (\ref{abssep100}) can be re-expressed as
\begin{eqnarray}
Tr[(\rho_{2\otimes 2}^{(iso)}(f))^{2}]&\leq& Max_{0\leq f\leq\frac{1}{2}}\frac{4f^{2}-2f+7}{18}
\nonumber\\&=& \frac{28}{72}
\label{trace22}
\end{eqnarray}
Therefore, the newly constructed balls $B_{1}$ and $B_{2}$ described by (\ref{abssep1}) and (\ref{trace200}) is bigger in size compared to the ball described by $Tr[(\rho_{2\otimes 2}^{(iso)}(f))^{2}]\leq \frac{1}{3}$ and hence the new ball contains more absolutely separable state. Also, we find that the ball $B_{2}$ contains more absolutely separable states than the ball $B_{1}$.
\subsubsection{Class of states in $2\otimes 3$ quantum system}
Let us consider a class of states in $2\otimes 3$ quantum system parameterized with two parameters $\alpha$ and $\gamma$, which is given by \cite{chi}
\begin{eqnarray}
&&\rho_{\alpha,\gamma}^{2\otimes 3}=\alpha(|02\rangle\langle 02|+|12\rangle\langle 12|)+\frac{4\gamma+2\alpha-1}{3} |\psi^{-}\rangle\langle \psi^{-}|+\nonumber\\&&
\frac{1-\gamma-2\alpha}{3}(|00\rangle\langle 00|+|01\rangle\langle 01|+|10\rangle\langle 10|+|11\rangle\langle 11|),~~\nonumber\\&&
0\leq \alpha \leq \frac{1}{2},~~ 0\leq \gamma \leq 1
\label{state23}
\end{eqnarray}
where $|\psi^{-}\rangle=\frac{1}{\sqrt{2}}(|01\rangle-|10\rangle)$. The state is separable if and only if $\alpha+\gamma\leq \frac{1}{2}$.\\
To simplify the calculation, let us choose $\gamma=\frac{1}{3}$. For this particular case, the state $\rho_{\alpha,\frac{1}{3}}^{2\otimes 3}$ is separable if and only if $0\leq \alpha\leq \frac{1}{6}$. Therefore, with this chosen value of $\gamma$, we can re-express the state $\rho_{\alpha,\frac{1}{3}}^{2\otimes 3}$ in terms of block matrices as
\begin{eqnarray}
\rho_{\alpha,\frac{1}{3}}^{2\otimes 3}=
\begin{pmatrix}
  X_{1} & Y_{1} \\
  Y_{1}^{\dagger} & Z_{1}
\end{pmatrix}
\end{eqnarray}
where $3\times 3$ block matrices $X_{1},Y_{1}$ and $Z_{1}$ are given by
\begin{eqnarray}
&&X_{1}=
\begin{pmatrix}
 \frac{2-6\alpha}{9}  &  0  & 0\\
   0 &  \frac{5-6\alpha}{18}  & 0\\
   0  &  0   &  \alpha
\end{pmatrix}, Y_{1}=\begin{pmatrix}
  0 & 0 & 0\\
  -\frac{1+6\alpha}{18} &  0  &  0\\
   0 &  0  &  0
\end{pmatrix},\nonumber\\&& Z_{1}=\begin{pmatrix}
  \frac{5-6\alpha}{18} & 0  & 0\\
   0 &  \frac{2-6\alpha}{9}  &  0\\
   0  &  0    &    \alpha
\end{pmatrix}, 0\leq \alpha \leq \frac{1}{6}
\end{eqnarray}
The eigenvalues of the state $\rho_{\alpha,\frac{1}{3}}^{2\otimes 3}$ arranged in descending order $(\varepsilon_{1}\geq \varepsilon_{2} \geq \varepsilon_{3}\geq \varepsilon_{4}\geq \varepsilon_{5}\geq \varepsilon_{6})$ for different ranges of $\alpha$ as \\
(i) When $0\leq \alpha \leq 0.134$
\begin{eqnarray}
\varepsilon_{1}=\frac{1}{3}, \varepsilon_{2}=\varepsilon_{3}=\varepsilon_{4}=\frac{2-6\alpha}{9}, \varepsilon_{5}=\varepsilon_{6}=\alpha
\label{eigenvsl23a}
\end{eqnarray}
(ii) When $0.134 \leq \alpha \leq \frac{1}{6}$
\begin{eqnarray}
\varepsilon_{1}=\frac{1}{3}, \varepsilon_{2}=\varepsilon_{3}=\alpha, \varepsilon_{4}=\varepsilon_{5}=\varepsilon_{6}=\frac{2-6\alpha}{9}
\label{eigenvsl23b}
\end{eqnarray}
It can be easily verified using (\ref{abssepcondgen}) that the state $\rho_{\alpha,\frac{1}{3}}^{2\otimes 3}$ represent absolute separable state for $0.019\leq \alpha \leq \frac{1}{6}$.\\
The ball $(B_{1})$ is described by\\
(i) When $0\leq \alpha\leq0.134$
\begin{eqnarray}
Tr[\rho_{\alpha,\frac{1}{3}}^{2\otimes 3}]^{2}\leq \frac{216\alpha^{2}-156\alpha+41}{9}
\label{b1i}
\end{eqnarray}
(ii) When $0.134\leq \alpha\leq \frac{1}{6}$
\begin{eqnarray}
Tr[\rho_{\alpha,\frac{1}{3}}^{2\otimes 3}]^{2}\leq \frac{1854\alpha^{2}-1452\alpha+377}{81}
\label{b1ii}
\end{eqnarray}
The ball $(B_{2})$ is described by
\begin{eqnarray}
Tr[\rho_{\alpha,\frac{1}{3}}^{2\otimes 3}]^{2}\leq \frac{2016\alpha^{2}-1488\alpha+389}{81},~~0\leq \alpha\leq\frac{1}{6}
\label{b2}
\end{eqnarray}
We also find in this example that the ball $(B_{2})$ contain more separable and absolute separable states than the ball $(B_{1})$.

\section{Absolute separability condition in terms of purity}
In this section, we will discuss the condition of absolute separability of the quantum state $\rho \in 2\otimes 2$ in terms of $Tr(\rho^{2})$. Then we generalize the absolute separability condition for the state belong to $H_{2}\otimes H_{d}$.\\
Let us consider a two-qubit state described by the density operator $\rho_{AB}^{2\otimes 2}$. The state $\rho$ would be absolutely separable if the purity of the state measured by $Tr[(\rho_{AB}^{2\otimes 2})^{2}]$ satisfies the inequality given in the following result.\\
\textbf{Result-5:} The state $\rho_{AB}^{2\otimes 2}$ is absolutely separable if and only if
\begin{eqnarray}
(\lambda_{1}-\lambda_{3})^{2}\leq 4\lambda_{2}Tr[(\rho_{AB}^{2\otimes 2})^{2}]\leq 4\lambda_{2}(\lambda_{3}+2\sqrt{\lambda_{2}\lambda_{4}})
\label{abscond22}
\end{eqnarray}
where $\lambda_{1}\geq \lambda_{2}\geq \lambda_{3} \geq \lambda_{4}$ denoting the eigenvalues of $\rho_{AB}^{2\otimes 2}$.\\
\textbf{Proof:} For $p>0$ and positive semi-definite matrix $\rho_{AB}^{2\otimes 2}$, we have
\begin{eqnarray}
\|(\rho_{AB}^{2\otimes 2})^{p}\|_{1}=\|\rho_{AB}^{2\otimes 2}\|_{p}^{p}
\label{norm22}
\end{eqnarray}
In particular, taking $p=2$ in (\ref{norm22}), we get
\begin{eqnarray}
&&\|(\rho_{AB}^{2\otimes 2})^{2}\|_{1}=\|\rho_{AB}^{2\otimes 2}\|_{2}^{2}\nonumber\\&\Rightarrow&
\sum_{i=1}^{4}\lambda_{i}[(\rho_{AB}^{2\otimes 2})^{2}]=\sum_{i=1}^{4}\lambda_{i}^{2}(\rho_{AB}^{2\otimes 2})
\nonumber\\&\Rightarrow& Tr[(\rho_{AB}^{2\otimes 2})^{2}]\leq \lambda_{1}(\rho_{AB}^{2\otimes 2})Tr(\rho_{AB}^{2\otimes 2})
\nonumber\\&\Rightarrow& Tr[(\rho_{AB}^{2\otimes 2})^{2}]\leq \lambda_{1}(\rho_{AB}^{2\otimes 2})
\label{norm221}
\end{eqnarray}
Using the absolute separability condition \cite{johnston}  in (\ref{norm221}), we get
\begin{eqnarray}
Tr[(\rho_{AB}^{2\otimes 2})^{2}]\leq \lambda_{3}+2\sqrt{\lambda_{2}\lambda_{4}}
\label{abscondupper}
\end{eqnarray}
Again we have
\begin{eqnarray}
&&Tr[(\rho_{AB}^{2\otimes 2})^{2}]=\sum_{i=1}^{4}\lambda_{i}^{2}(\rho_{AB}^{2\otimes 2})
\nonumber\\&\Rightarrow& Tr[(\rho_{AB}^{2\otimes 2})^{2}]\geq \lambda_{4}(\rho_{AB}^{2\otimes 2})Tr(\rho_{AB}^{2\otimes 2})
\nonumber\\&\Rightarrow& Tr[(\rho_{AB}^{2\otimes 2})^{2}]\geq \lambda_{4}(\rho_{AB}^{2\otimes 2})
\label{norm222}
\end{eqnarray}
Using the absolute separability condition \cite{johnston}  in (\ref{norm222}), we get
\begin{eqnarray}
Tr[(\rho_{AB}^{2\otimes 2})^{2}]\geq \frac{(\lambda_{1}-\lambda_{3})^{2} }{4\lambda_{2}}
\label{abscondlower}
\end{eqnarray}
Combining (\ref{abscondupper}) and (\ref{abscondlower}), we get (\ref{abscond22}). Hence proved.\\
We are now in a position to generalize the result-5 for qubit-qudit system.\\
\textbf{Result-6:} If the qubit-qudit state described by the density operator $\rho_{AB}^{2\otimes d}$ then it is an absolutely separable state if and only if
\begin{eqnarray}
(\lambda_{1}-\lambda_{2d-1})^{2}&\leq& 4\lambda_{2d-2} Tr[(\rho_{AB}^{2\otimes d})^{2}]\nonumber\\&\leq& 4\lambda_{2d-2}(\lambda_{2d-1}+2\sqrt{\lambda_{2d-2}\lambda_{2d}})
\label{abscond2d}
\end{eqnarray}
\textbf{Corollary-3:} The term in the R.H.S of the inequality (\ref{abscond2d}) is greater than or equal to the term in the R.H.S of the inequality (\ref{abssepcondgen}) i.e.
\begin{eqnarray}
4\lambda_{2d-2}(\lambda_{2d-1}+2\sqrt{\lambda_{2d-2}\lambda_{2d}}) \geq 4\lambda_{2d-2}\lambda_{2d}
\label{obs}
\end{eqnarray}
\textbf{Proof:} Let us recall the inequalities (\ref{abssepcondgen}) and (\ref{abscond2d}), which are re-expressed as
\begin{eqnarray}
(\lambda_{1}-\lambda_{2d-1})^{2}\leq 4\lambda_{2d-2}\lambda_{2d}
\label{obs1}
\end{eqnarray}
\begin{eqnarray}
(\lambda_{1}-\lambda_{2d-1})^{2}\leq 4\lambda_{2d-2}(\lambda_{2d-1}+2\sqrt{\lambda_{2d-2}\lambda_{2d}})
\label{obs2}
\end{eqnarray}
Now, consider the expression given as
\begin{eqnarray}
E=4\lambda_{2d-2}(\lambda_{2d-1}+2\sqrt{\lambda_{2d-2}\lambda_{2d}})-4\lambda_{2d-2}\lambda_{2d}
\label{obsexp}
\end{eqnarray}
Since $\lambda_{1} \geq \lambda_{2} \geq \lambda_{3} \geq ........\geq \lambda_{2d-2} \geq \lambda_{2d-1} \geq \lambda_{2d}$ so we have
\begin{eqnarray}
&& 4\lambda_{2d-2}(\lambda_{2d-1}+2\sqrt{\lambda_{2d-2}\lambda_{2d}})-4\lambda_{2d-2}\lambda_{2d}\nonumber\\&=&
4\lambda_{2d-2}(\lambda_{2d-1}-\lambda_{2d})+8\lambda_{2d-2}\sqrt{\lambda_{2d-2}\lambda_{2d}}\nonumber\\&\geq& 0
\label{obsexp}
\end{eqnarray}
Hence proved.\\
Therefore, corollary-3 shows that (\ref{abscond2d}) contains more separable and absolute separable states than the inequality given in  (\ref{abssepcondgen}).\\
Let us take an example of a state in $2\otimes 4$ dimensional system for which corollary-3 holds.\\
A quantum state in $2\otimes 4$ dimensional system described by the density operator $\sigma_{AB}^{2\otimes 4}$ is given by
\begin{eqnarray}
\sigma_{AB}^{2\otimes 4}=
\begin{pmatrix}
  \frac{1}{8} & 0 & 0 & 0 & \frac{1}{81} & 0 & 0 & \frac{1}{81} \\
  0 & \frac{1}{8} & 0 & 0 & 0 & 0 & 0 & 0 \\
  0 & 0 & \frac{1}{8} & 0 & 0 & 0 & 0 & 0 \\
  0 & 0 & 0 & \frac{1}{8} & \frac{1}{81} & 0 & 0 & \frac{1}{81} \\
  \frac{1}{81} & 0 & 0 & \frac{1}{81} & \frac{1}{8} & 0 & 0 & 0 \\
  0 & 0 & 0 & 0 & 0 & \frac{1}{8} & 0 & 0 \\
  0 & 0 & 0 & 0 & 0 & 0 & \frac{1}{8} & 0 \\
  \frac{1}{81} & 0 & 0 & \frac{1}{81} & 0 & 0 & 0 & \frac{1}{8} \\
\end{pmatrix}
\end{eqnarray}
The eigenvalues of $\sigma_{AB}^{2\otimes 4}$ are given by
\begin{eqnarray}
&& \lambda_{1}=\frac{97}{648}\geq \lambda_{2}= \lambda_{3} = \lambda_{4}= \lambda_{5} = \lambda_{6}= \lambda_{7} = \frac{1}{8} \geq \nonumber\\&& \lambda_{8}= \frac{65}{648}
\label{eig100}
\end{eqnarray}
For the state $\sigma_{AB}^{2\otimes 4}$, the inequalities (\ref{abssepcondgen}) and (\ref{abscond2d}) reduces to
\begin{eqnarray}
(\lambda_{1}-\lambda_{7})^{2}\leq 4\lambda_{6}\lambda_{8}
\label{obs11}
\end{eqnarray}
and
\begin{eqnarray}
(\lambda_{1}-\lambda_{7})^{2}\leq 4\lambda_{6}(\lambda_{7}+2\sqrt{\lambda_{6}\lambda_{8}})
\label{obs12}
\end{eqnarray}
It can be easily seen that the inequalities (\ref{obs11}) and (\ref{obs12}) are satisfied for the eigenvalues given in (\ref{eig100}). Thus,
the state $\sigma_{AB}^{2\otimes 4}$ is an absolutely separable state. Also, we find that the R.H.S of (\ref{obs12}) is greater than the R.H.S of (\ref{obs11}).

\textbf{Corollary-4:} If any qubit-qudit state violate the inequality (\ref{abscond2d}) then the qubit-qudit state under investigation is not absolutely separable.\\
To verify corollary-4, let us consider a $2\otimes 4$ dimensional state described by the density operator $\rho_{AB}^{2\otimes 4}$ which is given by \cite{ha} \begin{eqnarray}
\rho_{AB}^{2\otimes 4}=
\begin{pmatrix}
  \frac{71}{400} & 0 & 0 & \frac{7}{400} & 0 & 0 & \frac{7}{400} & 0 \\
  0 & \frac{39}{400} & 0 & 0 & \frac{71}{400} & 0 & 0 & \frac{23}{400} \\
  0 & 0 & \frac{31}{400} & 0 & 0 & \frac{31}{400} & 0 & 0 \\
  \frac{7}{400} & 0 & 0 & \frac{39}{400} & 0 & 0 & \frac{71}{400} & 0 \\
  0 & \frac{39}{400} & 0 & 0 & \frac{39}{400} & 0 & 0 & \frac{23}{400} \\
  0 & 0 & \frac{31}{400} & 0 & 0 & \frac{31}{400} & 0 & 0 \\
  \frac{7}{400} & 0 & 0 & \frac{39}{400} & 0 & 0 & \frac{39}{400} & 0 \\
  0 & \frac{23}{400} & 0 & 0 & \frac{23}{400} & 0 & 0 & \frac{111}{400} \\
\end{pmatrix}
\end{eqnarray}
The eigenvalues of $\rho_{AB}^{2\otimes 4}$ are given by
\begin{eqnarray}
&&\lambda_{1}=\frac{189+\sqrt{5321}}{800}\geq \lambda_{2}=\frac{17}{80}\geq \lambda_{3}=\frac{4}{25} \geq \lambda_{4}=\frac{31}{200}
\nonumber\\&\geq&\lambda_{5}=\frac{189-\sqrt{5321}}{800}\geq \lambda_{6}=\lambda_{7}=\lambda_{8}=0
\end{eqnarray}
Therefore, the state $\rho_{AB}^{2\otimes 4}$ violate the inequality (\ref{abscond2d}) and thus it is not an absolute separable state.\\
Further, the state $\rho_{AB}^{2\otimes 4}$ has been shown in \cite{ha} as separable state. Thus the state $\rho_{AB}^{2\otimes 4}$ is a separable state but not an absolute separable state.
\section{Conclusion}
To summarize, we have characterize the absolute separable states in terms of quantum correlation which can be measured by quantum discord. We found an instance of absolute separable states with such negligible amount of quantum correlation that can be approximated to zero but still it is useful in quantum algorithm to solve Deutsch-Jozsa problem. Since these absolute separable states have approximately zero quantum correlation so we expect that it can be prepared in the experiment easily and not only that these states give quantum advantage over classical with respect to the running time of the algorithms. This prompted us to investigate about the structure of the class of absolute separable states with zero discord. We found the class of absolute separable zero discord state which are residing within the ball described by $Tr(\rho^{2})\leq \frac{1}{3}$. Further, we find that there exist classes of absolute separable zero discord state that falls outside the ball. To fill this gap, we have derived new separability criterion, which we have used to construct a new ball that holds most of the absolute separable states lying in $2\otimes d$ dimensional Hilbert space. In particular, we have shown that the absolute separable states that lying outside the ball described by $Tr(\rho^{2})\leq \frac{1}{3}$, now residing inside the newly constructed ball. Thus, we conclude that the new ball is bigger in size and this fact is illustrated by giving few examples. The derived absolute separability condition in terms of purity may help in finding the upper and lower bound of the linear entropy of the absolute separable states. Since the bounds of the purity of absolute separable states can be expressed in terms of eigenvalues so it would be easier to estimate the bound of purity and hence linear entropy experimentally.

\end{document}